\documentclass[12pt]{article}
\usepackage{epsfig,float,latexsym}

\date{}
\begin{document}

\newcommand{\beq}{\begin{equation}}
\newcommand{\eeq}{\end{equation}}
\newcommand{\nn}{\nonumber}
\newcommand{\bea}{\begin{eqnarray}}
\newcommand{\eea}{\end{eqnarray}}

\title{Improving Learning in Science and Mathematics with Exploratory and Interactive Computational Modelling}

\author{$\mbox{Rui Gomes Neves}^1$\footnote{For correspondence: rgn@fct.unl.pt.},$\mbox{ Jorge Carvalho Silva}^2$ \&$\mbox{ V\'{\i}tor Duarte Teodoro}^1$\\
{\small \it $\mbox{}^1$ Unidade de Investiga\c {c}\~ao Educa\c {c}\~ao e Desenvolvimento (UIED)}\\
{\small \it Departamento de Ci\^encias Sociais Aplicadas, Faculdade de Ci\^encias e Tecnologia}\\
{\small \it Universidade Nova de Lisboa}\\
{\small \it Monte da Caparica, 2829-516 Caparica, Portugal}\\
{\small \it $\mbox{}^2$ Centro de F\'{\i}sica e Investiga\c {c}\~ao Tecnol\'ogica (CEFITEC)}\\
{\small \it Departamento de F\'{\i}sica, Faculdade de Ci\^encias e Tecnologia}\\
{\small \it Universidade Nova de Lisboa}\\
{\small \it Monte da Caparica, 2829-516 Caparica, Portugal}}

\maketitle

\begin{abstract}
Scientific research involves mathematical modelling in the context of an interactive balance between theory, experiment and computation. However, computational methods and tools are still far from being appropriately integrated in the high school and university curricula in science and mathematics. In this chapter, we present a new way to develop computational modelling learning activities in science and mathematics which may be fruitfully adopted by high school and university curricula. These activities may also be a valuable instrument for the professional development of teachers. Focusing on mathematical modelling in the context of physics, we describe a selection of exploratory and interactive computational modelling activities in introductory mechanics and discuss their impact on student learning of key physical and mathematical concepts in mechanics.
\end{abstract}

\newpage
\section{Introduction}

Science is an evolving structure of knowledge based on hypotheses and models which lead to theories whose explanations and predictions about the universe must be consistent with the results of systematic and reliable experiments (see, for example, \cite{Chalmers99}
-\cite{Feynman67}). The process of creating scientific knowledge is an interactive blend of individual and group reflections which involve modelling processes that balance theory, experiment and computation \cite{Blumetal07}
-\cite{Slootenetal06}. This cognitive frame of action has a strong mathematical character, since scientific reasoning embeds mathematical reasoning as scientific concepts and laws are represented by mathematical entities and relations. In this scientific research process, computational modelling plays a key role in the expansion of the cognitive horizon of both science and mathematics through enhanced calculation, exploration and visualization capabilities.       

Although intimately connected with real world phenomena, science and mathematics are thus based on abstract and subtle conceptual and methodological frameworks which change along far from straightforward evolution timelines. These cognitive features make science and mathematics difficult subjects to learn, to develop and to teach. In an approach to science and mathematics education meant to be effective and in phase with the rapid scientific and technological development, an early integration of computational modelling in learning environments which reflect the exploratory and interactive nature of modern scientific research is of crucial importance \cite{Ogborn94}. However, computers, computational methods and software, as well as exploratory and interactive learning environments, are still far from being appropriately integrated in the high school and university curricula in science and mathematics. As a consequence, these curricula are generally outdated and most tend to transmit to students a sense of detachment from the real world. These are contributing factors to the development of negative views about science and mathematics education, leading to an increase in student failure. 

Physics education is a good example to illustrate this situation. Consider the general physics courses taken by first year university students. These are courses which usually cover a large number of difficult physics topics following a traditional lecture plus laboratory instruction approach. Due to a lack of understanding of fundamental concepts in physics and mathematics, the number of students that fail on the course examinations is usually very high. Moreover, many students that eventually succeed also reveal several weaknesses in their understanding of elementary physics and mathematics \cite{HallounHestenes85}
-\cite{McDermottRedish99}. For example, in the Faculty of Sciences and Technology of the New Lisbon University, on average only less than 30 percent of the students are able to take such courses on the first attempt. Of these, less than 10 percent can be said to have acquired a solid knowledge of the taught general physics and associated mathematics topics.
      
Although it is clear that there are many reasons behind this problem, it is also clear that the solution has to involve changes in the physics education model. Indeed, many research studies have shown that the process of learning can effectively be enhanced when students are involved in the learning activities as scientists are involved in research activities \cite{McDermottRedish99}
-\cite{KeinerBurns10}. In addition, several attempts have been made to introduce computational modelling in research inspired learning environments. The starting emphasis was on professional programming languages such as Fortran \cite{Bork67} and Pascal \cite{RedishWilson93}. Although more recently this approach has evolved to Python \cite{ChabaySherwood08}, it still requires students to develop a working knowledge of programming, a generally time consuming and dispersive task which can hinder the process of learning physics. The same happens when using professional scientific computation software such as Mathematica and Matlab. To avoid overloading students with programming notions or syntax, and focus the learning process on the relevant physics and mathematics, several computer modelling systems were created, for example, Dynamical Modelling System \cite{Ogborn85}, Stella \cite{HPS97}, Easy Java Simulations \cite{ChristianEsquembre07} and Modellus \cite{Teodoro02}.

Besides being a curricular development problem, an adequate integration of computational modelling in the learning process of science and mathematics is thus also a technological problem. In this chapter, we discuss how Modellus (a freely available software tool created in Java which is able to run in all operating systems, see the software webpage at http://modellus.fct.unl.pt) can be used to develop computational modelling learning activities in science and mathematics with an exploratory and interactive character. These activities can be adopted by high school and university curricula. They may also be a valuable instrument for the professional development of teachers. Focusing on mathematical modelling in the context of physics, we describe examples of exploratory and interactive computational modelling activities on introductory mechanics which were implemented in a new course component of the general physics course taken by first year biomedical engineering students at the Faculty of Sciences and Technology of the New Lisbon University. Although conceived for a general physics course, these computational modelling activities are relevant for mathematics education as concrete applications of mathematical modelling \cite{Carson99}
-\cite{NRC89}. The activities were designed to emphasise cognitive conflicts in the understanding of physical concepts, the manipulation of multiple representations of mathematical models and the interplay between the analytical and numerical approaches applied to solve problems in physics and mathematics. As a domain general computer system for mathematical modelling \cite{Schwartz07}, Modellus is particularly well designed for this task because of the following main advantages: 1) an easy and intuitive creation of mathematical models using standard mathematical notation; 2) the possibility to create animations with interactive objects that have mathematical properties expressed in the model; 3) the simultaneous exploration of images, tables, graphs and object animations; 4) the computation and display of mathematical quantities obtained from the analysis of images and graphs.

\section{Course organization, methodology and student evaluation procedures}

Let us start by describing the educational stage where the computational modelling activities were implemented. The organization, methodology and evaluation strategies used here to introduce computational modelling in general physics can serve as a model to be adapted to other areas of science and to mathematics. 

The 2008/2009 general physics course for biomedical engineering involved a total of 115 students among which 59 were taking the course for the first time. Following the structure defined in the 2007/2008 edition \cite{Nevesetal09}, the course was divided into lectures, built around a set of key experiments where the general physics topics were first introduced, standard physics laboratories and the new computational modelling classes based on exploratory and interactive workshop activities.

In the computational modelling classes, the students were organized in groups of two or three, one group for each computer in the classroom. During each class, the student teams worked on a computational modelling activity set conceived to be an interactive and exploratory learning experience built around a small number of problems about challenging but easily observed physical phenomena. Examples are the motion of a swimmer in a river with a current or the motion of an airplane against the wind \cite{Nevesetal09}. The teams were instructed to analyse and discuss the problems on their own using the physical, mathematical and computational modelling guidelines provided by the activity documentation. To ensure adequate working rhythm with appropriate conceptual, analytical and computational understanding, the students were continuously accompanied and helped during the exploration of the activities. Whenever it was felt necessary, global class discussions were conducted to keep the pace and to clarify any doubts on concepts, reasoning and calculations. All computational modelling activities were created as interactive modelling experiments based on Modellus. In this course, each class activity consisted of a set of five modelling tasks in mechanics, presented in PDF documents, with text and embedded video support to help students both in class or at home in a collaborative online context centred on the Moodle online learning platform. In this course, the majority of the activity supporting text and videos presented complete step-by-step instructions to build the Modellus mathematical models, animations, graphs and tables. After constructing the models, students explored the multiple representations available to answer several questions about the proposed general physics problems. A few of the activities involved modelling problems where students saw only a video of the Modellus animations or graphs and then had to construct the corresponding mathematical models to reproduce them and answer the proposed questions.
      
The student evaluation procedures in the computational modelling classes involved both group evaluation and individual evaluation. For each computational modelling class, all student groups had to build five Modellus models, one for each task, and complete an online test, written in the Moodle platform, answering the questions of the corresponding activity PDF document. The individual evaluation consisted of the solution of two homework activities, each having three modelling problems to solve, and a test with two computational modelling problems. These activities involved problems based on those covered in the regular classes. The corresponding supporting text and videos gave only partial instructions to build the Modellus mathematical models, animations, graphs and tables. To answer the homework and test questions, students had to complete the mathematical models, animations, graphs and tables, partially revealed in the movies. To be evaluated, each student had to construct in Modellus one or two models for each problem, to be saved in Modellus files, and organize a Word document file with the answers to the proposed problems, illustrated with images of the Modellus mathematical models, animations, graphs and tables. Students unable to achieve a minimum grade of 40 in a 0-100 scale had to do an extra computational modelling examination, similar to the test. The final student classification was an average of the performance on the class activities, on the two homework activities and on the test (or exam) activities. All students also had to take pre-instruction and post-instruction Force Concept Inventory (FCI) tests \cite{Hestenesetal92} which did not count for their final grade. At the end of the semester, the students answered a questionnaire to access their degree of receptivity to this new computational modelling component of the general physics course.

\section{Computational modelling activities with Modellus}

For the computational modelling component of the general physics course we created workshop activities covering eight basic themes in mechanics \cite{Teodoro06,YF04}: 1) Vectors; 2) Motion and parametric equations; 3) Motion seen in moving frames; 4) Newton's equations: analytic and numerical solutions; 5) Circular motion and oscillations; 6) From free fall, to parachute fall and bungee-jumping; 7) Systems of particles, linear momentum and collisions and 8) Rigid bodies and rotations. Let us now discuss, as illustrative examples, two of the computational modelling activities about circular motion and oscillations, the theme opening the second part of the course. Again, we note that these are thought not only from the point of physics but also from a more traditional point of view of mathematics in order to help students make connections between different subjects.

A particle in circular motion (representing, for instance, a runner going around a circular track) describes a circle of radius $R$, a mathematical curve defined by $x^2+y^2=R^2$ in a Cartesian reference frame $Oxy$ whose origin is at the centre of the circle. In this frame, $x$ and $y$ are the Cartesian coordinates of the position vector $\vec{r}$. This vector has magnitude R and specifies where the particle is on the curve. As the particle moves around the circle, the magnitude $R$ is kept constant but the direction of the position vector changes with time. This direction is given by the angle $\theta$ that the position vector makes with the $Ox$ axis. The variables $R$ and $\theta$ define the polar coordinates of the position vector. The Cartesian coordinates $x$ and $y$ are also time dependent and are related to the polar coordinates $R$ and $\theta$ by trigonometric functions, $x=R\cos(\theta)$ and $y=R\sin(\theta)$.

\begin{figure}[H]
   \center{\psfig{file=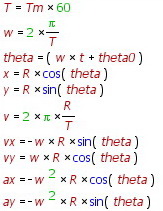,width=0.3\hsize}}
   \vspace{-0.2cm}
   \caption{Uniform circular motion: equations as seen in the Mathematical Model window of Modellus.}
   \label{fig1:modelcm}
   \end{figure}

To explore circular motion, students started with computational modelling activities about uniform circular motion. When the circular motion is uniform the particle traces one circle in every constant time interval $T$. This time interval is the period of the motion and its inverse $f=1/T$  is the frequency of the motion. The angle $\theta$ is then a linear parametric function of the time $t$, $\theta= \omega t+ \theta_0$  where $\omega=2\pi/T$ is the motion angular frequency, measured in radians per second, and $\theta_0$ is the initial direction of the position vector. The velocity $\vec{v}$ has a constant magnitude $v=\omega R$ and, being tangent to the circular trajectory of the particle, is always perpendicular to the position vector. The acceleration $\vec{a}$ has magnitude $a= \omega^2 R$ and a centripetal direction, that is, opposite to the direction of the position vector. The uniform circular motion is the composition of two simple harmonic oscillations, one along the $Ox$ axis and the other along the $Oy$ axis. These oscillations are characterized by the same amplitude $A=R$ and the same frequency $f=1/T$. The initial phase of the $Ox$ oscillation is $\theta_0$ and between them there is a time independent $\pi/2$ phase difference.

\begin{figure}[H]
   \center{\psfig{file=2ictma14circularmotion1phys.jpg,width=0.9\hsize}}
   \vspace{-0.2cm}
   \label{fig20:animationcm}
   \end{figure}

\begin{figure}[H]
   \center{\psfig{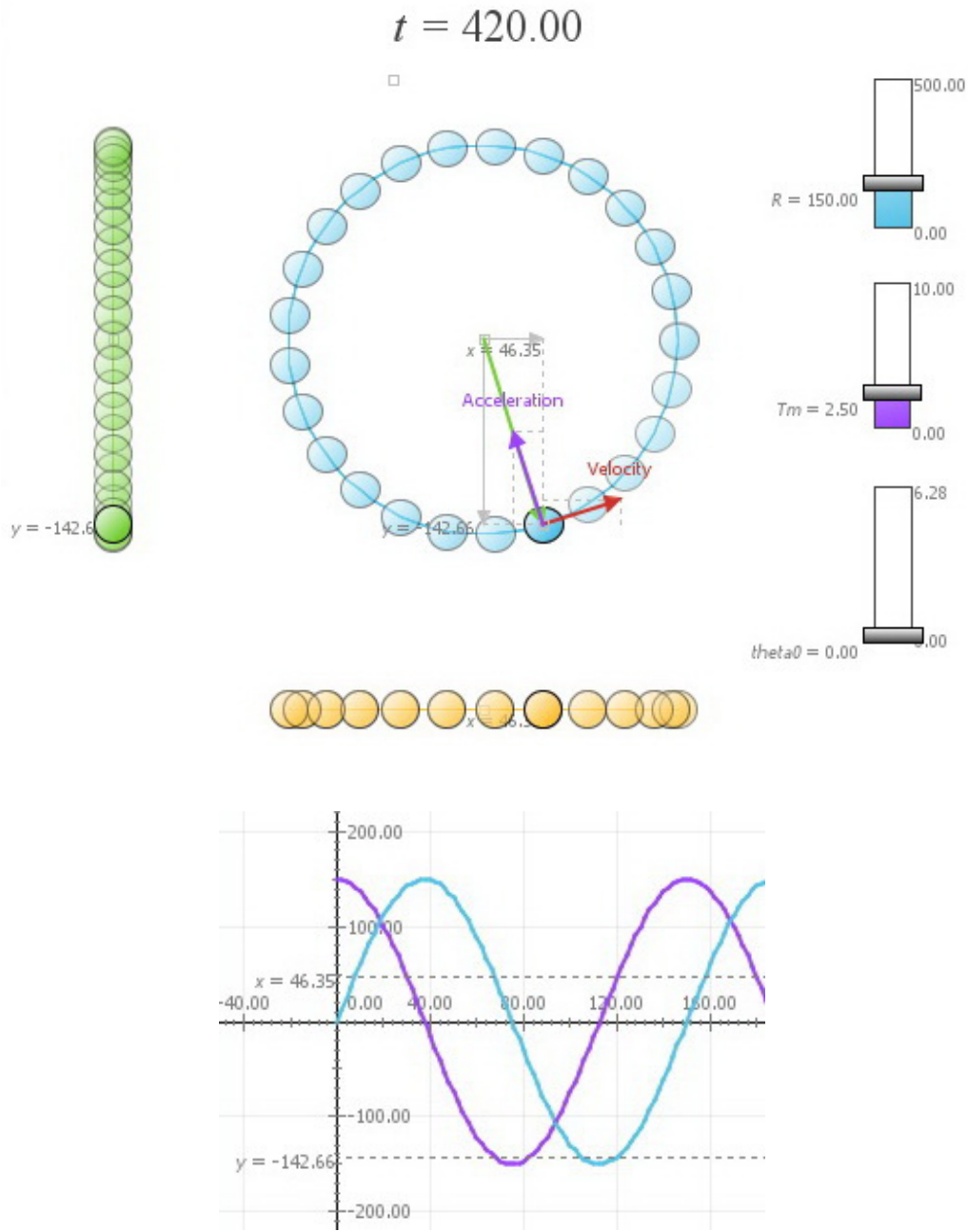}}
   \vspace{-0.2cm}
   \caption{Modellus animation and coordinate-time graphs for uniform circular motion.}
   \label{fig2:animationcm}
   \end{figure}
   
To model this type of motion, students had to recall what they learnt in the first part of the course during the computational modelling activities about vectors, parametric equations of motion, velocity and acceleration \cite{Nevesetal09}. Building upon this previously acquired knowledge, students were able to construct a model associating the Cartesian coordinates of the position vector to the corresponding polar trigonometric functions with the angle $\theta$ given by the linear parametric equation, $\theta= \omega t+ \theta_0$. They were also able to define the coordinates of the velocity and the acceleration (see Fig.~\ref{fig1:modelcm}). This mathematical model was complemented with graphs and tables of the different coordinate variables as functions of time, and by an animation allowing direct manipulation of the independent parameters of the model, $R,T,\theta_0$, as well as real time visual display of the trajectory of the moving particle, its position vector, velocity and acceleration. The harmonic oscillatory motions along the coordinate axis were also represented (see Fig.~\ref{fig2:animationcm}).

\begin{figure}[H]
   \center{\psfig{file=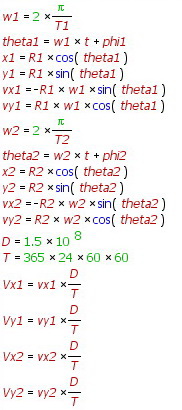,width=0.3\hsize}}
   \vspace{-0.2cm}
   \caption{Modellus mathematical model used by the students to solve the problem of the successive Earth and Mars oppositions.}
   \label{fig3:earthmars1}
   \end{figure}
   
With this model, students were able to explore, visualise and reify the initially abstract physical and mathematical concepts associated with uniform circular motion. For example, by combining the information from the several different simultaneous representations, they analysed the motion of a particle tracing a circle of radius $R=$ 150 m every 2.5 minutes, and were able to compare the velocity and the acceleration as functions of time and to calculate these vectors at time $t=$ 7 minutes.

During these computational modelling activities, students showed difficulties in distinguishing between a vector, like the velocity and the acceleration, and its magnitude. They were also puzzled when asked to solve the same problem considering the angles to be measured in degrees instead of radians. Indeed, at first students were frequently unable to create velocity and acceleration vectors with the right magnitude and direction. Similarly, they did not place the angle conversion factor in the correct place everywhere in the mathematical model. For example, in their first attempt they incorrectly multiplied the speed by $180/\pi$. To be able to correct the models and at the same time visualise the effect of the change in the animation and other model representations was for the students an essential advantage of the modelling process with Modellus in helping them to solve these learning difficulties.

\begin{figure}[H]
   \center{\psfig{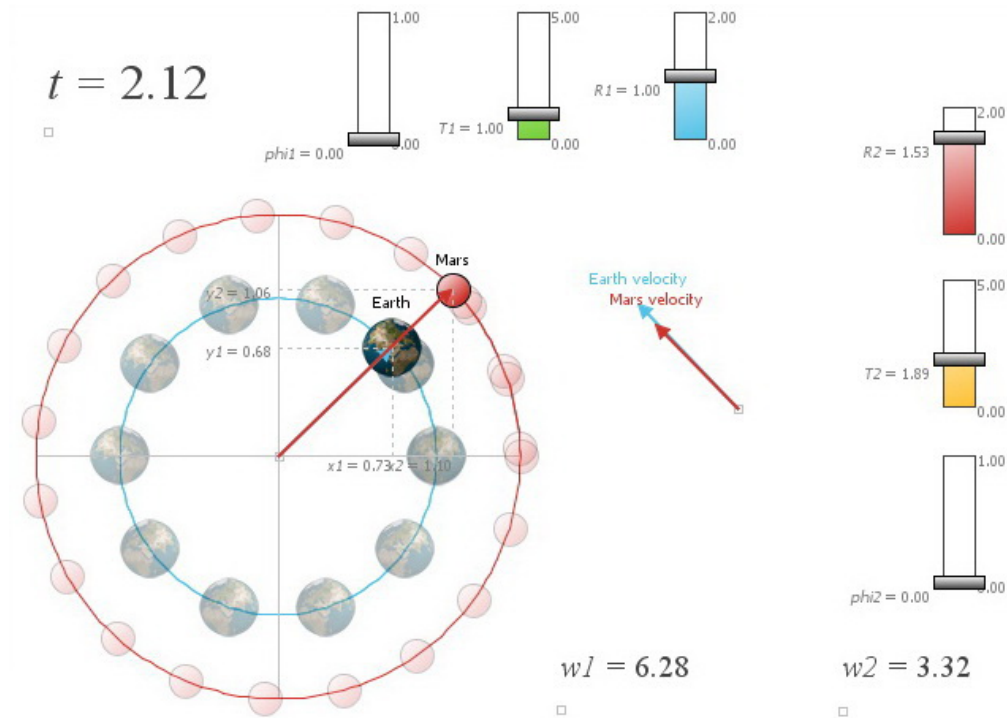}}
   \vspace{-0.2cm}
   \caption{A photograph of the Modellus animation at the moment of the first Earth and Mars opposition.}
   \label{fig4:earthmars2}
   \end{figure}

Using this trigonometric model, students were then capable to construct a model in Modellus to estimate the solution to the following astronomical problem: What is the time interval between two successive oppositions of the Earth and Mars? To help students, we suggested, in the activity PDF document, the assumption of considering the motions of the Earth and Mars around the Sun to be uniform circular motions. We also taught them to use the average Earth-Sun distance (known as the astronomical unit and denoted by AU) as the distance scale for the problem. In this scale, the average Earth-Sun distance is simply 1 AU and the average Mars-Sun distance is 1.53 AU. Taking into account that the approximate motion periods of the Earth and Mars are, respectively, 1 year and 1.89 years and using the year as the unit of time, students were able to develop a mathematical model (see Fig.~\ref{fig3:earthmars1}) and an animation (see Fig.~\ref{fig4:earthmars2}) representing the motions of the Earth and Mars around the Sun. In the process, they were able to determine the angular velocities of both planets and the time interval between two successive oppositions. Using the conversion factors 1 AU = 1.50$\times{10^8}$ km and 1 year = 3.15$\times{10^7}$ s, they were also able to find in km/s the orbital velocities of the Earth and Mars at the time of the model first occurring opposition (see Fig.~\ref{fig3:earthmars1}). To achieve the precision required by the Moodle online test, students used a position vector or velocity coincidence method (see Fig.~\ref{fig4:earthmars2}). The adjustment of the numerical step was an important numerical technique students learned to apply to obtain animations with realistic trajectories and correct answers to the questions of this astronomical challenge.

\section{Conclusions}
 
In this paper we have shown how Modellus can be used to develop exploratory and interactive computational modelling learning activities in science and mathematics. We have described a set of examples in introductory mechanics which were implemented in a course component of the general physics course taken by first year biomedical engineering students at the Faculty of Sciences and Technology of the New Lisbon University. We have shown that during class the computational modelling activities with Modellus were successful in identifying and resolving several student difficulties in key physical and mathematical concepts of the course. Of crucial importance in this process, was the possibility to have a real time visible correspondence between the animations with interactive objects and the object's mathematical properties defined in the model, and also the possibility of manipulating simultaneously several different representations like graphs and tables. We have also shown that with Modellus students can be exploring authors of models and animations, and not just simple browsers of computer simulations.

\begin{figure}[H]
   \center{\psfig{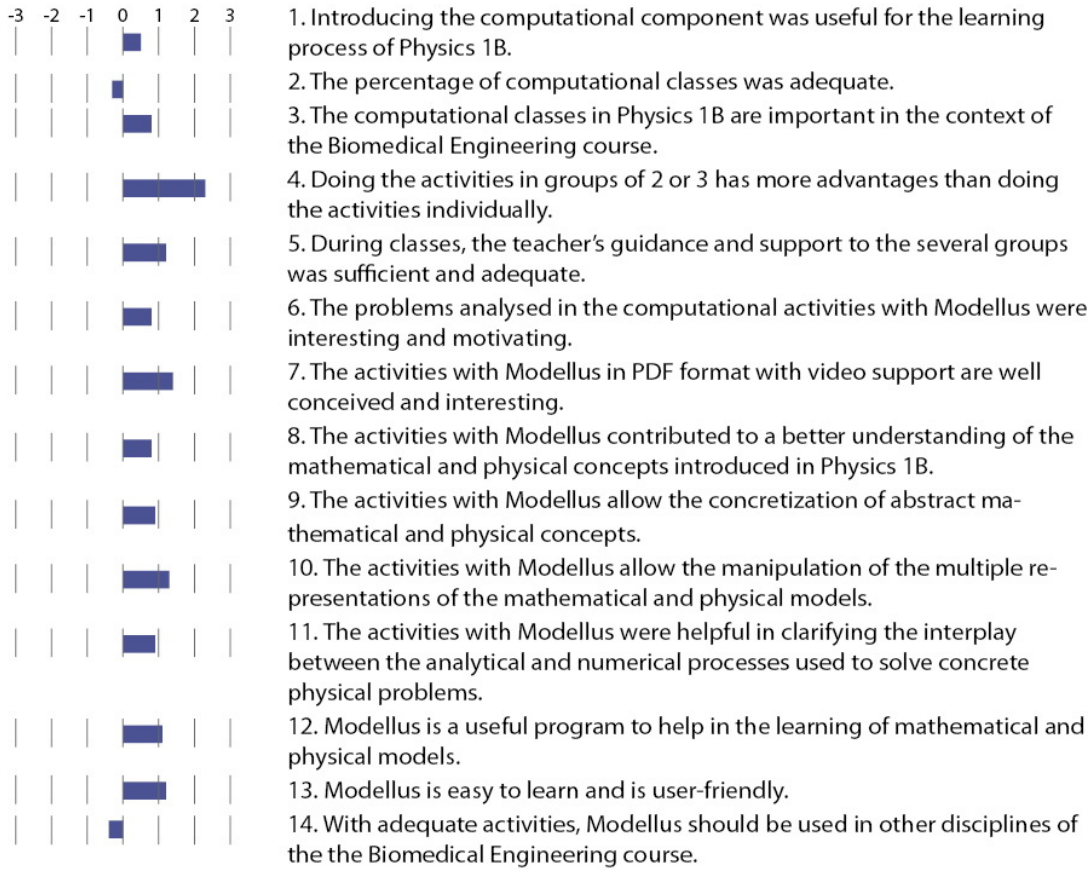}}
   \vspace{-0.2cm}
   \caption{Physics 1B questionnaire and results.}
   \label{fig5:results}
   \end{figure}

The successful class implementation of this set of computational modelling activities was reflected in the student answers to the questionnaire (see~\ref{fig5:results}) given at the end of the course. In this questionnaire students gave their opinion about a set of assertions characterizing the new Modellus computational modelling component, using a Likert scale starting at -3 and ending at +3, where -3 stated complete disagreement and +3 complete agreement. The remaining negative values stated partial disagreement and the remaining positive values partial agreement. The choice of the number 0 meant the student had no opinion about a particular statement. The 2009 results of the questionnaire, represented for each assertion by the average bar over all student answers, are shown graphically in Fig.~\ref{fig5:results}.

Globally, students reacted positively to the activities, considering them to be helpful in the learning process of mathematical and physical models. For them, Modellus was easy enough to learn and user-friendly. In this course, students showed a clear preference to work in teams in an interactive and exploratory learning environment. The computational modelling activities with Modellus presented in PDF documents with embedded video guidance were also considered to be interesting and well designed. A natural sense of caution in relation to novelty and to evaluation procedures was nevertheless detected. Students also felt that the content load was too heavy and that the available time spent on the computational modelling activities was insufficient.

In spite of global success during the class implementation phase, the FCI test results led to an average FCI gain of 22\%, an indication that the general physics course with the computational modelling component is just performing as a traditional instruction course (Hake, 1998). Although this performance score refers to the general physics course as a whole, the results of the questionnaire and students opinions about the computational modelling component also indicate that some aspects of the implementation approach should be changed. In this context, possible ways forward are: 1) Increase the relative importance and value of the computational modelling component; 2) Reduce the heavy content load (as perceived by students); 3) Increase time spent on the modelling tasks; 4) Choose problems more closely related with the specific subject of the student's major course; 5) Introduce less guided, more discovery oriented instruction guidelines as well as computational modelling problem finding.

\vspace{0.25cm}

\leftline{\large \bf Acknowledgements}
\vspace{0.25cm}

Work supported by Unidade de Investiga\c {c}\~ao Educa\c {c}\~ao e Desenvolvimento (UIED) and Funda\c {c}\~ao para a Ci\^encia e a Tecnologia (FCT), Programa Compromisso com a Ci\^encia, Ci\^encia 2007.

\end{document}